\documentclass[conference]{IEEEtran}
\usepackage{times,amssymb}
\usepackage{textcomp} %for \textbardbl
\usepackage{stmaryrd} %for \talloblong
\usepackage{graphicx}%for caption !
\usepackage{txfonts}%for \coloneqq etc
\usepackage{balance}
\usepackage{authblk}
\def\BibTeX{{\rm B\kern-.05em{\sc i\kern-.025em b}\kern-.08em
    T\kern-.1667em\lower.7ex\hbox{E}\kern-.125emX}}
\begin{document}
\newcommand {\pgm}[1]{Figure (\arabic{counter}). #1.\\ \addtocounter{counter}{1}}
\newcommand {\comment}[1]{\{#1\}\\}
\newcommand {\priority}[1]{{\bf sched\_priority(#1)}\\}
\newcounter{counter} 
\addtocounter{counter}{1}
\newcounter{subcounter}[counter] %though not used here but concept is simple and elegant
\newcommand {\subnum}{(\alpha{subcounter}). \addtocounter{subcounter}{1}}%probably not used
\newcommand{\mc}{Mobile Computing }
\renewcommand{\mu}{Mobile UNITY }
\newcommand{\fm}{formal methods }
\newcommand{\Fm}{Formal methods }
\newcommand{\FM}{Formal Methods }
\newcommand{\tab}{\hspace*{.6cm}}
\newcommand{\stab}{\hspace*{.4cm}}
\newcommand{\lt}{\textless}
\newcommand{\gt}{\textgreater}
\renewcommand{\null}{$\bot$}
\newcommand{\true}{{\it true}}
\newcommand{\false}{{\it false}}
\newcommand{\append}{$\bullet$}
\newcommand{\ceq}{$\coloneqq$}
\renewcommand{\neq}{$\not = $}
\newcommand{\acompsep}{$\textbardbl$} % assignment component separator, may be don't use \textbardbl at all
\newcommand{\sep}{$\talloblong$} %statement separator
\newcommand{\la}{$\langle$}
\newcommand{\ra}{$\rangle$}
\newcommand{\ens}{event notification service model}
\newcommand{\reactsto}{{\bf reacts-to }}
\newcommand{\IF}{\tab {\bf if }}
\newcommand{\commentout}[1]{}

\title {Formal Methods and Event Notification Systems in Mobile Computing Environment$^\ast$} 
\author[1]{Prashant Kumar}
\author[2]{R. K. Ghosh}
\affil[1,2]{Department of CSE, IIT Kanpur, INDIA\authorcr Email: {\tt \{prashankr, rkg\}@cse.iitk.ac.in}\vspace{1.5ex}}

\date{\today}
%\date{November 9, 2005}
%   \twocolumn[

%     \begin{@twocolumnfalse}
       \maketitle
	   \begin{abstract}
%	   {\hskip -0.5cm}
%		  {\bf write abstract here}

In this report, we have explored the issues associated with the 
specification of event based systems in a mobile environment using 
Unity~\cite{unity}. We used a few constructs and concepts from Mobile 
UNITY which was proposed as an extension of UNITY by Roman and 
McCann~\cite{intro}. Our aim in this report is to show that some of 
the constructs proposed in Mobile UNITY are not unnecessary. Those
constructs are overly powerful and put hindrance on the mapping
from UNITY specification to particular architectures, which 
is one of the key simplicity of UNITY specification. Using an example of 
a message based event notification system we have shown that a 
system with a simple modification to the structure of assign section
of the UNITY programs could serve
well in mappping and implementation at the same time preserve 
the small and compact proof logic of UNITY.

%	   In this paper we have discussed the constructs and program structure of UNITY. Then built upon then to conceptualize \mu . We have also discussed proof logic and formal details of various constructs.\\

%\noindent {\bf Keywords--} This and that.
\vspace{0.1in}

       \end{abstract}
%     \end{@twocolumnfalse}
%   ]

   {
     \renewcommand{\thefootnote}
       {\fnsymbol{footnote}}
	 \footnotetext[1]{
	As Part of CS697 : Special Topics in Computer Science}
%		 As part of CS697, under the guidance of Dr. R. K. Ghosh}
   }

%   {
%     \renewcommand{\thefootnote}
%       {\fnsymbol{footnote}}
%	 \footnotetext[2]{
%	As Part of CS697 : Special Topics in Computer Science}
%   }

%\input{intro}
\section{Introduction} \label{sec:intro}

\commentout{
Recent increase in availability of computing devices coupled with their 
compactness and widespread infrastructure supporting mobility  has led to 
the concept of ubiquitous computing. It has in some cases led to fundamental 
changes in which things have been done in traditional computing. For example, 
data dissemination, conventionally pull based systems have been the choice 
for various application in distributed computing domain. In pull based 
systems, client requests data from the server and after the request, server 
sends the data back to it, here  we note that it is client initiated  data 
delivery and the responsibility of data transfer rests with the client. 
Opposed to this in many applications, there has been a shift towards 
{\it push based} data transfer paradigm. It is a server centric approach 
to data transfer. Here the responsibility of data transfer lies with the 
server. Server transmits the data to to satisfy the needs of clients and  
clients have  to tune  their push clients thus enabling  them to receive only 
the information   in which they are interested.  {\bf write me}.
%	There are various facets it, one of them is also the emergence of push based information distribution. 
	Push based systems are more suited for the case where the nodes at the client side are low at computing resources and have limited communication needs. In real life applications one of the ways in which it is visible is in terms of feeds and notification wherein a client registers its interest with the server and then it is server's responsibility to send the message to the client whenever a message/event of interest is generated. In general these system may or may not support mobility but usability of these systems and convenience to users is further enhanced if they  can be support mobility. 
% By mobility we  mean {\bf write me}.
In fact mobility support seems to be a natural requirements for  event 
notification  systems, without  the support  for mobility their usefulness 
will greatly reduce.  Moreover due to asynchronous and loose coupling 
characteristics of push based notification systems support for mobility 
is not only desirable rather it is in some sense natural for these systems 
to support mobility.
 But then there are many issues which come up  with mobile agents and 
mobile computing environment. One has to take into account that environment 
may be fragile and the nodes may have limited resources.	
% {\bf write me: blend formal methods here}
}

The purpose of this work is to investigate the usefulness of formal 
methods, particularly UNITY~\cite{unity} and Mobile UNITY~\cite{intro}, in 
specification and reasoning about applications for mobile computing 
environments.
%event notification systems particularly for mobile 
A Mobile Computing Environment (MCE) can be considered as a asynchronous 
Distributed Computing Environment (DCE). The major difference between MCE
and DCE which affect semantics of developing correct program seems to be in
modeling failures. In DCE a failure to communicate is considered as a 
failure of the system as a whole. But in MCE, failures to communicate may be
transient as mobile terminals frequently change their locations.
Another notable difference originates from the intended usage
scenarios of mobile devices. It leads to demand for context dependent 
services from the applications running on a mobile computing environment. 
There are variety of reasons other than changing location for a 
participating node in an MCE to lose connectivity with rest of the 
network~\cite{satyanarayanan2001pervasive}.

We are particularly interested in evaluating the UNITY
model as our tool for formal specification. UNITY was introduced by Misra 
and Chandy~\cite{unity} in late 80s. It is a formal computational
model and with a proof system that supports development of correct
programs and their efficient implementation on asynchronous parallel/
distributed architectures. Subsequently, some intensive research 
centered around UNITY. It includes  modifications by Misra~\cite{misra1988general,progress} , 
Dynamic UNITY~\cite{du}, Context UNITY~\cite{roman2007modeling} and  Mobile UNITY~\cite{intro}.
Yet, the original UNITY still remains a seminal work in formal 
computational model and proof systems.

The focus of our work was to evaluate Mobile UNITY vis-a-vis UNITY in 
building mobile applications. The main purpose of Mobile UNITY was to
develop a new notation and an underlying formal model for "supporting
specification and proof system for decoupled, location-aware 
systems"~\cite{du}. Essentially it provides a formal layer on the top of 
UNITY's concurrency model. Therefore, we concentrated on the technical 
modifications to UNITY that have been proposed in Mobile UNITY to find 
if such modifications are necessary. 

% This encouraged us to examine the usefulness of mobile 
%UNITY with respect to ordinary UNITY.

In order to evaluate Mobile UNITY against UNITY, we considered 
the implementation of the Event Notification System (ENS) on mobile 
computing environment. The reasons for choosing ENS implementation
in mobile computing environment as a case study can be explained as follows.
In any distributed computing model, a computation is represented
by a sequence of local computation followed communication between 
compute nodes. So, message passing or the underlying communication 
pattern is fundamental to interaction of computational tasks. 
There are two main modes of communication, namely, 
\begin{enumerate}
    \item pair-wise communication, and
    \item multiple participants communication.
\end{enumerate}
Each mode of communication is further classified as asymmetric 
(client-server), or symmetric communication (peer-to-peer). 
Communication with multiple participants can be unicast 
(one-to-one), multicast (one-to-many) or convergecast (many-to-one).  
An ENS scheme generalizes these communication pattern to 
$m$-to-$n$ scheme, where $m, n\ge 1$. Developing
ENS in mobile computing environment~\cite{arnold2005method} further generalizes 
this many-to-many communication pattern. Obviously, 
mobility support can be considered as a natural requirements for event
notification systems in mobile computing environment.
% without the support  for mobility their usefulness will greatly
%reduce. 
Moreover, push based data delivery model is found to be eminently
suitable for data dissemination in mobile computing environment~\cite{acharya1995dissemination}.
Due to asynchronous, loose coupling characteristics of push based 
notification systems the support for mobility in some is sense natural for 
these systems. 

Among the additional notations which were proposed in Mobile UNITY for
modification of UNITY, the most powerful construct is the \reactsto. 
But this construct is found to be most difficult from point of view of 
implementation. Furthermore, it also complicates the proof system of 
original UNITY. 
%proposed some changes to Mobile UNITY and shown that one of its key constructs 
%can  of the construct introduced by Mobile UNITY were not ideally suited for 
%implementation purpose.
%We have proposed minimal addition to UNITY model and 
%shown these modifications are sufficient for implementation of event 
%based systems in mobile environment. 
In this paper we propose only minimal modifications to {\bf assign}
section of UNITY while preserving its weakly fair execution model 
for supporting specification and reasoning about implementing ENS in mobile 
environment. The proposed modifications to UNITY are designed
in a way that they do not complicate the proof system of the UNITY model
like \reactsto construct of Mobile UNITY. We also prove that \reactsto 
is not needed, and %and using minimal 
%modificiations of the kind suggested in this paper, 
UNITY without support of any such highly powerful constructs can still be  
sufficient for specification of event based mobile applications like ENS
by using minimal modification of the kind suggested in this paper.
%Consquently, the  can be dropped.
Since the proposed changes do not violate the execution 
model and semantics of UNITY, it preserves the proof system as well. 
%		\subsection{Motivation}
%\subsection{Contributions}

Contributions of this work can be summarized as follows:  
\begin{itemize}
\item  Simplifying the notations in Mobile UNITY framework  for 
specification of systems in Mobile environment.
\item Accompanying simplification in the proof part for various properties 
of the system.
\item Restructuring the {\bf assign} section of UNITY program to make it closer to real world systems from the point of view of implementation.
\item Showing that some powerful constructs of Mobile UNITY eliminated and
still it can support both specification and reasoning about decoupled, 
location aware systems.
\item To the best of our knowledge ours is the first use of UNITY 
    formalism for message based event notification service systems.
\end{itemize}

%\subsection{Paper Structure}
The rest of the paper is organised as follows. Section~\ref{sec:ens} 
gives an overview of event notification systems. 
Sections~\ref{sec:unityall} 
and~\ref{sec:munity} provide an overview of UNITY and Mobile UNITY
respectively. In section\ref{sec:events} we discuss events in UNITY and 
Mobile UNITY and towards the end of this section we also describe the
additions to UNITY which though necessarily minimal help in
specification and implementation of system for real world systems in 
mobile environment. It is followed by a detailed discussion of formal
specification of event notification system and 
its implementation in section~\ref{sec:application}. Finally 
section~\ref{sec:conclusion} summarises our contribution and lists the 
direction for future work.
\section{Event Notification Service Systems} \label{sec:ens}
Communication between nodes has always been a central issue 
in distributed computing. Newer issues have appeared in communication
over wide area network in the context of loosely coupled systems and 
mobile agents. Such distributed computing environments do not 
guarantee low latency or continuous and reliable connectivity as
one finds in a wired network of distributed system. So it has led 
to newer abstractions like event based interaction for design and 
implementation of software systems for such environments. 
Some works related to event notification service system such as 
SIENA~\cite{siena}, JEDI~\cite{huck1998jedi} and many others support this view.
%SIENA\cite{siena} , a research project aiming to build a generic {\it scalable event notification service} is based on similar assumption of event based systems.
Event notification systems are meant to disseminate information based on 
occurrence of some events. It involves a set of event producer and 
a set of event consumers.  Typically, the   
agents interested to receive notifications subscribe to a server and register themselves as the 
consumer of the events. The server dispatches the events to the subscribing agents
whenever a something of interest occurs.
	 
%	Event notification systems should be scalable and the software technology used to implement these must be light, more so if they are to be used in Mobile Computing environment. 
To motivate the point that such systems are needed and have there place in 
the larger scheme of things in design of applications, we compare message 
based event notification systems to traditional client-server model.
%and also with middlewares  based on distributed objects .
We also discuss the need for message based communication abstraction
based on events compared to distributed objects. 

%Here it is instructive to compare event notification service systems with traditional client server computing model.
We can think of a client-server computing model to implement \ens where 
servers are producer and clients are consumer. 	Client-server model has
the advantage of being conceptually simple and has been  been quite 
effective for point to point communication model over the time. But 
there is a fundamental difference, the client-server model is 
{\it synchronous} and communication typically is  one-to-one whereas 
an event notification service systems provide both {\it time} and 
{\it space} decoupling. Time decoupling means  
interacting parties need not be present at the same time, and 
space decoupling means that the interacting parties need not be 
co-located. Apart from this there are several point to point 
communication middleware such as CORBA, RMI and DCOM which can 
offer these services but they are primarily meant for synchronous 
invocation of some service  offered by a remote server and normally 
would be heavyweight and overkill if the task of the application is 
more basic and primitive. Hence to address the issues related to the
systems where communication needs are not elaborate, other alternatives
have been explored. Emergence of message based systems is one such
solution. Before discussing further we note that message based 
communication is not a new concept in distributed computing but the 
lightweight communication needs have made them the most suitable choice.
%We briefly  discuss about messages in the following paragraph and again 
%compare them with distributed objects. 

A message is essentially a structured piece of information sent from one 
agent to another over a communication channel.  Messages may deliver data, 
meta data (such as acknowledgements) or notification  to an agent. 
Messaging services  have a pre-defined and commonly agreed protocol between 
the agents. Message passing is not as robust and sophisticated as 
distributed objects and is necessarily a restrictive communication 
scheme compared to object based communication schemes like CORBA and RMI 
or  even with respect to RPC. However, they are simple and elegant solution for 
the situations where communication needs dictated by application are 
limited and minimal. If seen in wider perspective, the goal of these two
are quite different. Distributed objects serve the purpose of extending the 
application across the network and normally provide a local handle to 
the remote object for calling its methods. On the other hand, message passing is meant 
for a simpler role. It defines a bare minimum protocol for sending data 
and also avoids overhead associated with most distributed object 
technologies. To summarize we note that  message passing based systems
are desired in following scenarios\cite{dcj}:
\begin{enumerate}
\item Communication needs are relatively simple in nature. 
\item Transaction throughput is critical. 
\item Scope of the system is limited.
\end{enumerate}

Before discussing more about event based notification system we turn 
our focus on formal methods. It is central to explain 
the key contributions of this paper.

\section{UNITY Formalism}
\label{sec:unityall}
We start with a brief introduction to UNITY. An interested reader may refer 
to~\cite{unity} for further details.
UNITY is a computational model and a proof system. It provides a 
minimal set notations to denote programs. UNITY programs can be seen as 
{\em a program in unbounded non-deterministic iterative transformation 
notation}. Some of these terms will become more clear in course of the
discussion below.

\subsection{UNITY Program Structure}
Following schema describes the structure of a UNITY program::
\label{subs:ustr}

{\it program} $\longrightarrow$

\begin{tabbing}
%=====================\=================\= ======================\= \kill
=============\===============\= \kill
%Zero\>first \>second\>third\\
{\bf Program}      \>  {\it program-name}   \\
{\bf declare}      \>  {\it declare-section}   \\
{\bf always}       \>  {\it always-section}   \\
{\bf initially}    \>  {\it initially-section}   \\
{\bf assign}       \>  {\it assign-section}   \\
{\bf end}
\end{tabbing}

\noindent We illustrate some of these features by a simple program {\it semaphore}.

\noindent {\bf Program} {\it semaphore}\\
{\bf declare} \tab   g : integer\\
{\bf initially}\tab  g = 1\\
{\bf assign }\\
\tab \la \sep $i : 0 \leq$ N :: \\
\tab\tab g,p[i] := g-1,false if b $>$0 $\wedge$ p[i]\\
\tab \sep g,v[i] := g+1,false if v[i]\\
\tab \ra\\
{\bf end}\{semaphore\}\\
{\centering \pgm{semaphore}}

%Now we explain the UNITY notation using the program.
%Few comments about this program as well as about notation of UNITY 
%are in order.

The variable names with their types are provided
in {\it declare-section}. Normally, boolean and integers are used as 
basic types. Arrays and sets or any other ADT can also be used.
The above program for semaphore does not use {\bf always} section.
But this section is for defining certain variables as function of others.
It is, therefore, equivalent to \#define of conventional 
C programming languages. The initial value of some of the variables are 
defined by {\bf initially} section. The uninitialized variables can 
assume arbitrary values from its domain. The {\bf assign} section 
consists of a set of assignment statements is the heart of UNITY program.
The program execution starts in the state where values of the variables are 
as specified in the {\bf initially} section. Then these statements are 
executed atomically and non-deterministically in 
weakly fair manner. By {\it weakly fair} we mean that in an 
infinite computation each statement is scheduled for execution {\em infinitely 
often}. The symbol \sep~acts as separator between statements and 
\la~\ra~is used to limit the scope of quantification variables. One more 
point to be noted is that UNITY programs have no input/output statements.
All I/O is assumed to be performed by appending items to or removing items 
from the corresponding sequences.  
\commentout{
In the program {\it semaphore}, a semaphore {\it g} 
is shared between $N$ processes. The $i^{th}$ process, 0 $\leq i\lt N$, 
requests a {\bf p}-operation on the semaphore by setting the boolean 
variable {\tt p[i]} to {\it true}. This {\bf p}-operation can be executed 
only if  $g\gt 0$.  The result of execution {\bf p}-operation is that 
{\tt p[i]} becomes {\it false} and $g$ is decremented by 1. Similarly, 
the $i^{th}$ process requests a {\bf v}-operation on the semaphore by 
setting {\tt v[i]} to {\it true}.  On completion of this operation {\tt v[i]}
is to be set to {\it false} and $g$ incremented by 1. The declarations of
{\tt p[i]}s and {\tt v[i]}s and setting their initial value to {\it true} 
is done in the respective {\tt p[i]}s. 
}
Here, we do not discuss the composition models of individual 
programs in UNITY, i.e., {\it union} and {\it superposition}. The readers
may refer to~\cite{unity} for the same. It may, however, be noted in 
passing that in standard UNITY, variables with same name are shared 
between programs. Other constructs from UNITY will be introduced as we
come across them in our discussion. 

\subsection{Execution Model}
In a UNITY program, more than one statement may be enabled at a time.
Concurrency is modeled by weakly fair interleaved execution of these 
atomic statements. %operations.
As stated earlier {\em weakly fair} means that in an 
infinite computation each statement is scheduled for execution infinitely 
often. Interestingly, a very simple notation like weakly fair
execution model has been used by UNITY to specify and verify correctness
large software systems in industry~\cite{inds} as well as for specification
of %the solution of the
many well known problems in distributed systems~\cite{Liu2018}.% Together 
%with this , the proof logic of UNITY though being equally minimal can be
%applied in formal derivation and verification of such programs. 

\subsection{Proof Logic in UNITY}
\label{subsec:prooflogic}
In the UNITY proof logic and program properties are expressed using a 
small set of predicate relations whose validity can be derived directly 
from the program text or from other properties through the application of 
inference rules.
%A proof of correctness is the demonstration that the 
%text of a program meets a certain conditions. 
System properties can be 
categorized in two fundamental groups: {\it safety} and {\it liveliness} 
properties. The authors in~\cite{def} have shown that any property  of the 
system can be expressed as the {\it intersection} of a pure safety and a 
pure liveliness property. Liveliness property essentially means that during 
the execution, program makes some progress and consequently some 
desirable state eventually does occur and safety property underlines that 
some {\it unsafe} state can not materialize during the execution of the 
program. 

The safety property has been described using {\bf constrains} relation. For 
brevity it is written as {\bf co} henceforth~\cite{gulwani2008program}. This 
construct is not unique to UNITY and can be used for any state system. 
Given two state predicates $p$ and $q$  the expression $p$ {\bf co} $q$
means that any state satisfying $p$, the next state in the execution must
satisfy $q$. For example, to express that some variable {\tt counter} in 
the program is monotonically increasing, we write 
\begin{center}
counter \ceq  $n$ {\bf co} counter $\geq n$ 
\end{center}
As an aside we note that in terms of {\it Hoare triple}~\cite{hoare69},
{\bf co} can be denoted as
\begin{center}
$p$ {\bf co} $q \equiv$ \la $\forall s::\{p\}s\{q\}$ \ra
\end{center}
To be precise, we should have quantified $n$ with universal quantifier 
in the above {\bf co} relation, but for convenience we will leave that part 
whenever it is clear from the context. We can build on more complex safety 
properties using {\bf co}. For example, {\bf invariant} that a variable 
always between 0 and $N$, i.e.,
\begin{center} 
{\bf invariant} 0 $\leq$ var $\leq N$
\end{center}
needs to be broken in two parts. 
In first part we have to verify from the {\bf initially-section} 
that initial value of var satisfies this property. Then using {\bf co} 
property on program text we can  assert that invariant indeed holds.

Now let us consider liveliness property. Here we implicitly use UNITY's 
inherent fairness assumption. Progress/liveliness is expressed in 
UNITY using the {\bf transient} relation where {\bf transient} $p$, 
state that the predicate $p$ is eventually falsified, i.e.
\begin{center}
{\bf transient} $p \equiv$ \la $\exists s :: \{p\}s\{\neg p\}$  \ra
\end{center}
which denotes the existence of a statement which when executed in 
a state satisfying $p$, produces a state that does not satisfy $p$.
 Using {\bf co} and {\bf transient}  we can construct the familiar 
operator {\bf ensures}. The relation $p$ {\bf ensures} $q$ means 
that for any state satisfying $p$ and not $q$, the next state must
satisfy $p$ or $q$. In addition, there is some statement $s$ that 
guarantees 
the establishment of $q$ if executed in a state satisfying $p$ and not $q$.
Because of fairness assumption, we can guarantee that this statement 
will eventually be selected for execution
\begin{center}
p {\bf ensures} q $\equiv$ (p $\wedge$ $\neg$q {\bf co} p $\vee$ q) $\wedge$ transient(p $\wedge$ $\neg$q)
\end{center}
%		\section{Mobile UNITY}
\section{Mobile UNITY}
\label{sec:munity}
\mu is an extension of UNITY with the aim of modeling systems in mobile 
environment as against UNITY which presents itself as a powerful
approach to concurrency for distributed systems but its focus is essentially 
on the static nature of computation. % that can be expressed.

\subsection{Adding Constructs to UNITY to make it \mu}
We now discuss some new constructs proposed by~\cite{intro} which
when added to UNITY make it suitable for specifying and reasoning 
about programs suited for mobile computing environment.
\commentout{ We discuss 
about {\it location} of an agent, {\it interaction} between various 
{\it components} of a section in a UNITY program, and in 
{\it interaction} section we describe the new type 
of statements, namely, {\it {reactive, inhibitive}} and {\it transactions} 
as proposed by~\cite{intro}.
}
The new constructs of \mu are illustrated with the example motivated 
by~\cite{intro}. For more details, the readers may refer to~\cite{intro}
and~\cite{comp}. System description is as follows. There are {\tt m} 
producers and {\tt n} consumers {\it all}
in a mobile environment ({\tt n} may or may not be equal 
to {\tt m}). %Producers are producing some information for clients. 
%We ill try to formalize a solution for this problem.
We take a mix of top down and bottom up approach to formalize the
solution to the problem. We build the whole system incrementally, 
but describe smaller parts top down. First of all we consider two 
standard UNITY programs for {\it sender} and {\it receiver}
\newline
\newline
{\bf Program}   {\it sender}\\
{\bf declare} \\
\tab bit : boolean\\
\sep \stab word: array[0..N-1] boolean\\
\sep \stab csend,crecv : integer\\
{\bf initially}\\
\tab  bit= 0\\
\tab  csend= N\\
{\bf assign }\\
\tab transmit::bit,csend\ceq word[csend],csend+1\\
\tab {\bf if} csend $<$ N $\wedge$ csend =  crecv\\
\sep \stab new :: word, csend \ceq NewWord(), 0 \\
\tab {\bf if} csend $\geq$ N\\
{\bf end}\\
\pgm{{\it sender}}
\newline
\newline
{\bf Program} {\it receiver}\\
{\bf declare} \\
\tab bit : boolean\\
\sep \stab buffer : array[0..N-1] boolean\\
\sep \stab csend,crecv : integer\\
{\bf initially}\\
\tab  bit= 0\\
\sep \stab crecv = N \\
{\bf assign }\\
\tab receive :: buffer[crecv+1], crecv \ceq\\
\tab  bit, crecv+1 {\bf if} crecv $<$ N $\wedge$ crecv \neq csend\\
\sep \stab reset :: crecv \ceq -1\\
\tab {\bf if } crecv $\geq$ N $\wedge$ csend = 0\\
{\bf end}\\
\pgm{{\it receiver}}

These are two standard UNITY programs where variables with same name 
are shared. We also note that UNITY as well in \mu comments are enclosed 
in \{\}. The first major and fundamental difference which comes in 
mobile environment is that we can not have shared name space. Here we 
assume that variables in different programs are not the same even if 
they have same name. So a variable {\tt v} in program {\tt P} is {\tt P.v}
but we may write it simply as {\tt v} in cases where there is no ambiguity. 
With these comments we give tentative structure of this system in \mu. 
\newline
\newline
{\bf System} {\it sender-receiver}\\
\tab {\bf program} {\it sender(i)}  {\bf at $\lambda$}\\
\tab \dots\\
\tab {\bf end}\\
\tab {\bf program} {\it receiver(i)}  {\bf at $\lambda$}\\
\tab  \dots\\
\tab {\bf end}\\
{\bf Components}\\
\tab receiver(0)  {\bf at $\lambda_0$}\\
\sep \stab sender(1) {\bf at $\lambda_0$}\\
{\bf Interactions}\\
\dots\\
{\bf end}\\
\pgm{ Tentative structure for {\it sender-receiver} system}
Now let us examine this program, taking up new concepts one by one.

\subsection{Location Awareness}
First of all let us consider notion of location of mobile agents. \mu has 
a special variable denoted by $\lambda$, which is mandatory in all programs.
% Conventionally it is denoted by {\bf $\lambda$}. 
\mu places no constraints on the type of $\lambda$. Intuitively, it should 
represent the notion of location of mobile agent/device and changes in
value of $\lambda$ denotes mobility. From \mu's point of view whether 
$\lambda$'s value is one dimensional or multidimensional (latitude /longitude 
IP address in network or memory address) doesn't make 
much difference. But from implementation point of view, we assume that agent 
has some way to update the value of $\lambda$

\subsection{Interaction between Mobile Agents}
Interaction section of the Mobile UNITY program helps in specifying the 
interaction between various agents/components. Its need basically arises 
because of distinct address spaces of agents in Mobile UNITY. 

\noindent{\bf Program} {\it sender(i) }{\bf at $\lambda$}\\
{\bf declare} \\
\tab bit : boolean\\
\sep \stab word : array[0..N-1] boolean\\
\sep \stab c : integer\\
{\bf initially}\\
\tab $\lambda$ =  SenderLocation(i)\\
{\bf assign }\\
\tab transmit::bit,c\ceq word[c],c+1 {\bf if} c $<$ N \\
\sep \stab new :: word,c \ceq NewWord(),0 {\bf if} c $\geq$ N\\
{\bf end}\\
\pgm{ Modified Version of {\it sender}}
\newline
\newline
{\bf Program} {\it receiver(j)}{\bf at $\lambda$}\\
{\bf declare} \\
\tab bit : boolean\\
\sep \stab buffer : array[0..N-1] boolean\\
\sep \stab c : integer\\
{\bf assign }\\
\tab zero :: c \ceq 0 \reactsto bit =  1 $\wedge$ c $\geq$ N\\ 
\sep \stab receive ::buffer[c],c\ceq bit,c+1 {\bf if} c $<$ N \\
\sep \stab move:: $\lambda$ \ceq buffer \\
\tab \reactsto validLoc(buffer) $\wedge$ c $\geq$ N\\
{\bf end}\\
\pgm{{\it Modified Version of receiver}}
We assumed that first bit of sender is 1 and added mobility to 
receiver that on receiving a word which represents valid location it moves 
to that position.

Now all we need to do is fill in {\bf Interactions} of sender-receiver system 
as follows: 

\noindent{\bf Interactions}\\
\tab receiver(j).bit \ceq sender(i).bit \\
\tab \reactsto sender(i).$\lambda$ = receiver(j).$\lambda$\\
\sep \stab {\bf inhibit} sender(i).transmit {\bf when}\\
\tab sender(i).c $>$ receiver(j).c $\wedge$\\
\tab sender(i).$\lambda$ =  receiver(j).$\lambda$\\
\sep \stab {\bf inhibit} receiver(j).receive {\bf when}\\
\tab receiver(j).c $\geq$ sender(i).c $\wedge$\\
\tab sender(i).$\lambda$ =  receiver(j).$\lambda$\\
%\subsubsection{Additional Constructs Needed}
%{\it New Constructs }:
Mobile UNITY introduces three new constructs:
{\bf inhibition}, {\bf transaction} and {\bf reaction}. We discuss these
constructs one by one.

\subsubsection{Inhibitions}
Inhibitions provide mechanism to constrain the non deterministic 
scheduler in certain undesirable state. It is particularly useful in global 
context, i.e., in interaction of agents. Its syntax is as follows.
\begin{center}
{\bf inhibit} label {\bf when } predicate 
\end{center}
We note following things regarding its semantics
\begin{itemize}
\item The key word {\bf when} has meaning similar to {\bf if}, but it is
conventionally used in inhibitions for emphasis. 
\item So, the above {\bf inhibit} statement can actually be rewritten
as %construct is same as\\ statement 
\begin{center}
{\bf if} $\neg$predicate $<$statement labeled label$>$ 
\end{center}
\end{itemize}
\subsubsection{Transactions}
A transaction is a sequence of statements enclosed in angle brackets 
and separated by semi colon. In other words, all statements within
brackets should be executed in the order they appear with no other
statement to be scheduled in between. Reactive statements are the only 
exception which can be triggered during the execution of transaction 
as well. Its syntax is as follows
\begin{center}
label:: $<$statement\_1; ...; statement\_n$>$
\end{center}
Transaction may or may not be inhibited.

\subsubsection{Reactions}
Reactive statements are similar to {\it exception} with one exception 
that after reactions control flow goes to the position where it was before 
reaction was triggered. Its syntax it 
\begin{center}
{\it assignment statement} \reactsto {\it predicate}
\end{center}
It is triggered whenever {\it predicate} is {\true}. The imporatant points
about reactions are the following:
\begin{itemize}
\item Set of reactive statements execute till the {\it fixed point} of this 
set reached.
\item Reactive statements must not be inhibited.
\item Reactive statements can trigger {\it during} the transaction.
\end{itemize}

The semantic requirements of reaction construct are not
only too strong but also it hinders efficient implementation. For example, 
a method suggested by the authors was that reactive set be checked after 
the execution of each statement. Moreover, as can be seen in the 
paper~\cite{intro} that it complicates the proofs also.
During specification, it is the responsibility of the programmer
to take care that reactions reach the fixed point otherwise it
may violate UNITY execution model, since even in infinite executions 
only reactions can keep executing. This problem can be taken care of by 
ensuring that in all scenarios reactions reach the fixed point but in
real life systems ensuring this may not be easy.

\subsubsection{Modifications to Proof Logic}
Mobile UNITY also suggests some modification to proof logic. It requires 
that statement $s$ in proof be replaced by $s*$ which is a transaction and 
may contain reactive statements as well. The interested reader 
may refer to~\cite{comp} for further details.

\section{Events in UNITY and Mobile UNITY} \label{sec:events}
Though the constructs introduced by Mobile UNITY, make the specification 
simple, the key construct \reactsto is too powerful and is difficult to 
implement. The authors of Mobile UNITY have suggested that after the 
execution of each statement, reactive set be checked. Clearly from point
of view of implementation % opinion this
the suggested approach is too naive.  In fact, the authors of~\cite{intro}
share this view too.
%	but they argue that ...... . 

The actual advantage of \reactsto is that it can asynchronously trigger the 
execution of some statement but the implementation overhead for the same 
as indicated is prohibitively high. We have adopted a new approach to the 
tackle the need for having \reactsto construct. 
We propose spliting of the {\bf assign} section in two parts with different 
scheduling priorities, but enforce that within each block weakly fair 
execution semantics of UNITY should be honored. We also note that based 
on application's requirements 
{\bf assign} section can be divided into arbitrary number of blocks. 
But for the concreteness of discussion, we just consider two blocks.
In fact, in most of the systems two blocks will suffice.
We also note that two different transactions were also proposed in  
Dynamic UNITY\cite{du} where their primary goal was to use UNITY for 
specification of dynamically changing systems. Now we discuss the 
interpretation  of these changes from the  proof logic point of view of 
UNITY.
From the proof logic point of view these parts of {\bf assign} 
are same before. Statements within these will be treated in exactly the 
similar manner to what they would have been had the statements been in 
the assign section. But from the execution model point of view  and 
particularly from the scheduling and implementation point of view, there is 
some difference. We propose that, there be different scheduling priority of 
the statements in these sections. We must take care of not violating the 
{\it weakly fair} execution semantics of UNITY, hence scheduling should be 
such that each section has a {\it non zero } finite probability of being 
selected over a finite period of time. In fact, statements in {\bf assign} 
section can be grouped in more than two levels, but the scheduling should 
satisfy the {\it weakly fair } execution semantics. A simple policy 
for assigning different scheduling priority to different transition 
sections could be say $1/3$ and   $2/3$  in case of two sections, greater 
value indicates higher priority.

\section{Implementing Event Notification System} %Applications Discussed}
\label{sec:application}
To validate the utility of the modification to UNITY suggested in this paper,
we examine how Event Notification System can be implemented using UNITY 
with the modified constructs. We are mainly concerned with the view of an
application which is sufficient to  demonstrate our 
modification and at the same giving a reasonably stand-alone view of the 
application. Thus we consider the structure of the application to be 
as follows.
\begin{itemize}
\item It has multiple servers as well as multiple clients. 
\item Mobility is modelled by migration of client(s) from one server to 
another. 
\item The applications models fragile environment in the form of 
seemingly abrupt disconnections and takes care that messages are not 
lost in the process.
\end{itemize}
%	View of the application: Event Notification System and focus on the fundamental concepts in message passing over here. 
Our focus over here is to show that such systems can be 
conveniently expressed in UNITY without using \reactsto and our proposed 
modifications can lead to effective implementation and using UNITY proof logic 
we can assert about some of the properties of the system. 

\subsection{UNITY specification}
\noindent {\bf Program} {\it client}(i) {\bf at $\lambda$}\\
{\bf declare} \\
$\lambda$ : location\\
\sep \stab interface: queue of message\\
\sep \stab server\_addr: address\\
\sep \stab in,out: queue of message\\
\sep \stab registered,subscribed: bool\\
{\bf always} \\
server $\equiv$ server\_addr\\ 
\sep \stab self $\equiv$ self\_addr\\ 
{\bf initially}\\
\comment{\null  is $ null$}
interface =  \null\\
server\_addr = \null\\
{\bf assign }\\
\priority{P1}
 \comment{update location}
 $\lambda$ \ceq update($\lambda$)\\
\comment{receive messages from interface}
\sep \stab 
interface,in \ceq tail(interface),in\append head(interface)\\
\IF interface \neq \null $\wedge$ head(interface).destination = self \\
\tab $\wedge$ head(interface).type = M\\
\comment{transmit messages from interface}
\sep \stab 
head(interface).status\ceq  send(head(interface))\\
\IF interface \neq \null $\wedge$ $\neg$head(interface).status  \\
\comment{registering with a new server: handoff}
\sep \stab 
interface\ceq interface\append \\
\tab msg(false,self\_addr,new\_server,H,(msg\_stats,server\_addr))\\
\IF registered $\wedge$new\_server \neq server\_addr\\
\comment{remove N messages from interface}
\comment{it shows that some request was not satisfied}
\sep \stab interface \ceq tail(interface)\\
\IF interface\neq \null $\wedge$ head(interface).msg=N\\
\priority{P2}
\comment{clear sent message from interface}
\sep \stab interface\ceq tail(interface)\\
\IF interface \neq \null $\wedge$ head(interface).status  \\
\comment{register}
\sep \stab 
interface\ceq \\
\tab interface\append msg(false,self\_addr,server\_addr,R,content) \\
\IF $\neg$ registered $\wedge$ can\_send(self,server)\\
\comment{set the registered flag}
\sep \stab 
registered,interface\ceq true,tail(interface)\\
\IF head.type= R $\wedge$head.msg=Y\\
\comment{subscribe}
\sep \stab 
out\ceq out\append(msg(false,self\_addr,server\_addr,S,content)) \\
\IF $\neg$subscribed\\
\comment{set the subscribed flag}
\sep \stab 
subscribed,interface\ceq true,tail(interface)\\
\IF head.type=S$\wedge$head.msg=Y\\
\comment{unsubscribe}
\sep \stab 
out\ceq out\append(msg(false,self\_addr,server\_addr,U,content)) \\
\IF unsubscribe\\
\comment{reset the unsubscribe flag}
\sep \stab 
unsubscribe,interface\ceq false,tail(interface)\\
\IF head.type=U$\wedge$head.msg=Y\\
\comment{update subscription}
\sep \stab 
out\ceq out\append(msg(false,self\_addr,server\_addr,P,content))\\
\IF subscribed$\wedge$update\_add\\
\sep \stab 
out\ceq out\append(msg(false,self\_addr,server\_addr,Q,content))\\
\IF subscribed$\wedge$update\_del\\
\comment{reset the update flag}
\sep \stab update,interface\ceq false,tail(interface)\\
\IF head.type=P$\wedge$head.msg=Y\\
\comment{reset the update\_del flag}
\sep \stab 
update\_del,interface\ceq false,tail(interface)\\
\IF head.type=Q$\wedge$head.msg=Y\\
\comment{de-register: voluntary}
\sep \stab 
out\ceq out\append(msg(false,self\_addr,server\_addr,D,content))\\
\IF deregister \\
\comment{reset de-registered flag}
\sep \stab 
deregister,interface\ceq false,tail(interface)\\
\IF head.type=D$\wedge$head.msg=Y\\
\comment{transfer messages from out queue to interface}
\sep \stab  interface,out\ceq interface\append head(out),tail(out)\\
\IF out \neq \null \\
\comment{handle messages on in queue}
\sep \stab 
tag[head(in).msg\_type],in\ceq head(in).tag,tail(in)\\
{\bf end\\}
\pgm{{\it client}}
{\bf Program} {\it server\_}(i){\bf at $\lambda$}\\
{\bf declare} \\
$\lambda$: location\\
\sep \stab subscr: array[num\_message\_type] of queue of clients
\sep \stab interface: queue of message\\
{\bf always} \\
self $\equiv$ self\_addr\\ 
{\bf initially}\\
interface =  \null\\
{\bf assign }\\
\priority{P1}
\comment{register clients}
registered\_clients\ceq \\
\tab registered\_clients \append client\\
\IF client$\notin$ registered\_clients\\
\comment{de-registering clients}
\sep \stab 
registered\_clients\ceq delete(registered\_clients, client) \\
\comment{transmit messages from interface}
\sep \stab \la \sep $i :0\leq i <$ length(interface)::\\
at(interface,i).status= send(at(interface,i)) \\
\IF interface \neq \null $\wedge$ $\neg$at(interface,i).status \ra \\
\priority{P2}
\comment{update subscription}
\sep \stab  subscr[msg\_type]\ceq arr[msg\_type]\append client \\
\IF head.type= P $\wedge$ client$\notin$arr[msg\_type]\\
\sep \stab
subscr[msg\_type]\ceq delete(arr[msg\_type],client) \\
\IF head.type= Q $\wedge$ client$\in$arr[msg\_type]\\
\comment{transfer messages from out queue to interface}
\sep \stab  interface,out\ceq interface\append head(out),tail(out)\\
\IF out \neq \null \\
\comment{make sent messages null on interface}
\sep \stab \la \sep i :0$\leq i < $length(interface)::\\
at(interface,i) \ceq \null \\
\IF interface \neq \null $\wedge$ at(interface,i).status \ra \\
\comment{remove null messages from head of interface}
\sep \stab interface\ceq tail(interface)\\
\IF head(interface)= \null\\
%\comment{append  messages to out queue } not needed 
%\comment{send request for old messages} later?
%\comment{sending info about client which moved} later?
{\bf end\\}
\pgm{{\it server\_}(i,addr\_self){\bf at $\lambda$}}

{\bf System}{\it ens-system}\\
{\bf Components}\\
\la\sep $i : 0 \leq i \le$ NumClients ::client($i$) at $\lambda_i$\ra\\
\sep \stab \la\sep $j : 0 \leq j  \le$ NumServers ::server($j$) at $\lambda_j$\ra\\
{\bf Interactions}\\
\la \sep $i,j: 0  \leq i\le$NumServers$\wedge$ 0$\leq$j$\le$NumClients::\\
\tab client($i$).new\_server\ceq server($j$)\\
\IF can\_send(server(i),client(j))\ra\\
{\bf end\\}
\begin{table}
	\begin{center}
	\begin{tabular}{|c|l|}
		\hline
		Message  type
		&
		Stands for\\
		\hline

		M
		&
Normal message \\
		\hline

H
		&
Registering with new server\\
		\hline
R
		&
Register\\
		\hline
S
		&
Subscribe\\
		\hline
U
		&
		Unsubscribe\\
		\hline

P
		&
		Update subscription(add)\\
		\hline
Q
		&
		Update subscription(delete)\\
		\hline

D
		&
		De-register\\
		\hline

	\end{tabular}
\caption{Message Types} 
\label{tbl:msg}
	\end{center}
\end{table}

\subsection{Discussion}

The above specification does not use \reactsto. In general, any program 
containing \reactsto can be converted to one without having \reactsto. 
A simple heuristic is to replace \reactsto by {\bf if} and put all these in 
higher priority section of the program. Now we explain some of the subtle 
points of the above specification. It is assumed that $P1>P2$, 
in client as well as in server.  We have used a 
function $can_send(i,j)$ which returns if agent $i$ can send message to 
agent $j$. Both client and server maintain $in$ and $out$ queues, which are 
queues of messages. The messages on $in$ queue come from the $interface$ and 
messages from  $out$ queue go to $interface$.  There is a 
difference in the way messages are transmitted from the $interface$ in case of 
$client$ and $server$. In case of a client, the message at the head of the 
queue is processed, it will block if  the first message in the queue can not 
be sent.  Whereas this is not the case with server. The rationale
for it is that at a time server may have connection with one client but
may not have with another and if the client for which message is at the front 
of the queue is not connected at a time, all the other clients which can 
receive message should be able to 
receive. That is why transmission of message from the interface of server 
works on all the messages in $interface$ simultaneously. Various message 
types have been explained in Table~\ref{tbl:msg}.
New server actually comes at a lower level with hand off, that is why 
we have not shown it explicitly over here.  The specification of handoff would 
be needed had we been working at level of Mobile IP~\cite{} and 
Cellular IP~\cite{}. We have 
not shown the handling of messages on in queue of client and addition of 
messages to out queue of server, because these are application dependent 
decisions, and all we need here and assume is that clients handle the 
message they get on their in queue in a way consistent with its application 
logic, similarly out of server too has messages for clients consistent with 
the associated application logic.

\section{Conclusion}
\label{sec:conclusion}
	In this work we have shown that UNITY formalism is quite powerful for specification of  systems in mobile computing environment. Based on our experience with Mobile UNITY we have proposed doing away with very strong Mobile UNITY construct \reactsto. We have also proposed modification to the {\bf assign} section of UNITY to reflect the systems in Mobile computing. We have shown that our additions do not violate the weakly fair execution semantics of UNITY model. These modifications do not modify the proof logic system  of UNITY.  We have taken  an example of event notification service system to illustrate our ideas more concretely. As a future work one could try to incorporate these changes to the compilers which exist for translations of UNITY program to conventional programming languages.
%appendix
%\input{ack}
\balance
\bibliographystyle{ieeetr}

\end{document}